\documentclass[12pt]{article}

\usepackage[colorlinks=false]{hyperref}
\usepackage[dvips]{graphicx}
\usepackage[usenames,dvipsnames]{color}
\usepackage{epsfig}
\usepackage{pdfpages}
\usepackage{rotating}
\usepackage{amssymb}
\usepackage{amsmath,amsfonts}
\usepackage[latin1]{inputenc}
\usepackage{verbatim}
\usepackage[tableposition=top,font=small,labelfont=bf,format=hang]{caption}
\usepackage{booktabs}
\usepackage{color}
\newtheorem{lem}{Lemma}
\newtheorem{thm}{Theorem}
\newtheorem{prop}{Proposition}

\newcommand{\F}{\mathbb{F}}
\newcommand{\Z}{\mathbb{Z}}
\newcommand{\Q}{\mathbb{Q}}

\newcommand{\vc}{{\bf c}}

\newenvironment{proof}{\noindent\textbf{Proof.}\quad}
{\hspace{\stretch{1}}%
\rule{1ex}{1ex}\\}

\begin{document}

\title{ Construction of isodual codes \\from \\polycirculant matrices\thanks{This research is supported by
National Natural Science Foundation of China (61672036),
Excellent Youth Foundation of Natural Science Foundation of Anhui Province (1808085J20), Academic fund for outstanding talents in universities (gxbjZD03)}
}

\author{
Minjia Shi, Li Xu\thanks{Minjia Shi and Li Xu, School of Mathematical Sciences, Anhui University, Hefei, Anhui, 230601,
China, {\tt smjwcl.good@163.com, 382039930@qq.com}}, 
Patrick Sol\'e\thanks{CNRS, University of Aix Marseille, Centrale Marseille, I2M, Marseille,  France, {\tt sole@enst.fr}}
}

\date{}
\maketitle

\begin{abstract}
Double polycirculant codes are introduced here as a generalization of double circulant codes. 
When the matrix of the polyshift is a companion matrix of a trinomial, we show that such a code is isodual, hence formally self-dual. Numerical examples show that the codes constructed have optimal or quasi-optimal parameters amongst formally self-dual codes. Self-duality, the trivial case of isoduality, can only occur over $ \F_2$ in the double circulant case.
Building on an explicit infinite sequence of irreducible trinomials over $\F_2,$ we show that binary double polycirculant codes are  asymptotically good.

\end{abstract}


{\bf AMS Classification (MSC 2010):} Primary 94B05, Secondary 11C08
\section{Introduction}

Self-dual codes is one of the most fascinating class of codes as witnessed by their many connections with codes, lattices, and block designs \cite{MS}. This class has been enlarged, in recent years, to {\em isodual} codes that is to say codes that are equivalent to their duals \cite{AAHS,B,KL,MS}. These constitute in turn a subclass of the older class of {\it formally self-dual codes} that is to say codes the weight enumerator of which is a fixed point of the MacWilliams transform \cite{BH,DGH1,DGH2,FPGH}. A very popular and successful construction technique for isodual codes is the use of circulant matrices. In particular double circulant codes are easily shown to be isodual \cite{BGH}. In the present paper, we generalize this technique
to polycirculant matrices, and introduce double polycirculant codes.

In \cite{ADLS,LPS} was studied the notion of polycyclic codes, that is linear codes over a finite field $F,$ that are invariant under a generalized shift (called here a {\em polyshift}),
and affording a structure
of ideal over a ring of the form $R_f=F[x]/\langle f\rangle$ for some $f \in F[x]$ (the case $f=x^n-1$ is that of classical cyclic codes). While the name was coined in \cite{LPS}, the concept (under the name pseudo-cyclic code) has been known for a long time
\cite{PW}. As is well known, polycyclic codes are shortened cyclic codes, and conversely shortened cyclic codes are polycyclic \cite[p.241]{PW}. 

In the present paper, we introduce and study a class of codes called double polycirculant codes (DP) from the standpoint of duality, minimum distance, and asymptotic performance.
A matrix is called {\em polycirculant} if its rows are successive polyshifts of its first row. A DP code is then a linear code with generator matrix of the form $(I,A)$ where $I$ is an identity
matrix, and $A$ a polycirculant matrix. Thus DP codes reduce to double circulant codes when the polyshift is the classical shift.
When $A^t=QAQ$ for $Q$ a permutation matrix it is easy to show that the DP code is equivalent to its dual, and is, in particular, formally self-dual (FSD). FSD codes have
been studied extensively over $\F_2$ \cite{BH,FGPH,KP},  $\F_3$ \cite{DGH0}, $\F_4$ \cite{HaK}, and even $\F_5,$ or $\F_7$ \cite{DGH}. Indeed, we can show that DP codes can be binary self-dual only if they are double circulant.
We focus on the special case when the matrix of the polyshift is the companion matrix of a trinomial with nonzero constant term. In that situation, every polycirculant matrix satisfies the condition on its transpose
mentioned above. For $q=2,3,5,7$ numerical examples in short to medium lengths show the DP codes have parameters equal or up to one unit of the best-known FSD codes. Further, by random coding, we can show that the relative distances of long binary codes in that family satisfy a family of lower bounds that is, up to epsilons, the Gilbert-Varshamov bound for linear codes of rate one half. They thus constitute
a class of asymptotically good codes. The argument relies on the construction of an infinite family of irreducible trinomials, a fact of independent interest, and  on some properties of cyclic vectors in linear algebra \cite[chap. 7]{HK}. This generalizes the asymptotic properties of self-dual
double-circulant codes \cite{AOS}, and of self-dual negacirculant codes \cite{AGOSS,SQS}. The proof is restricted to $\F_2$ where we can show that an infinite family of irreducible trinomials exists. 

The material is arranged as follows. Section 2 collects the notations and notions needed to follow the rest of the paper. Section 3 studies duality properties of DP codes: isoduality, self-duality and evenness.
Section 4 derives asymptotic bounds by random coding. Section 5 displays some numerical examples of parameters of DP codes. Section 6 concludes the article and points out some significant and challenging open problems.
\section{Preliminaries}
\subsection{Linear codes}
Throughout, $F$ denotes a finite field.
If $q$ is a prime power, we denote by $\F_q$ the finite field of order $q.$ Let $N$ denote a nonnegative integer.
A {\bf (linear) code} $C$ of length $N$  over a finite field $\mathbb{F}_{q}$ is a $\mathbb{F}_{q}$ vector subspace of  $\mathbb{F}_{q}^N.$ The dimension of the code is its dimension as a $\mathbb{F}_{q}$ vector space, and is denoted by $k.$ The elements of $C$ are called {\bf codewords}. Two codes $C$ and $D$ are {\bf equivalent} if there is a monomial matrix $M$ such that $MC=D$  (recall that a matrix is {\bf monomial} if it contains exactly one nonzero element per row and per column; in particular any permutation matrix is a monomial matrix).
The {\bf (Hamming) weight}  of $x \in \F_q^N$ is denoted by $W(x).$  The minimum nonzero weight $d$ of a linear code is called the {\bf minimum distance}.
The {\bf dual} $C^ \bot$ of a code $C$ is understood w.r.t. the standard inner product. A code is {\bf self-dual } if it is equal to its dual, and {\bf isodual } if it is equivalent to its dual.
A code is {\bf formally self-dual} (FSD) if it has the same weight distribution as its dual. Thus isodual codes are FSD.
A binary code is {\bf even} if the weights of all its codewords are even, and {\bf odd} otherwise.
If $C(N)$ is a sequence of codes of parameters $[N, k_N, d_N]$, the {\bf rate} $r$ and {\bf relative distance} $\delta$ are defined as $$r=\limsup\limits_{N \rightarrow \infty}\frac{k_N}{N} \ {\rm and} \
\delta=\liminf\limits_{N \rightarrow \infty}\frac{d_N}{N}.$$
A family of codes is said to be {\bf good} if it contains a sequence with rate and relative distance such that $r\delta >0.$
Recall the classical {\bf entropy function} $H(x)$ of the real variable $x$ defined for $0<x<1,$ by the formula
$$H(x)=-x\log(x)-(1-x)\log(1-x).$$
See \cite[Chap. 10, \S 11]{MS} for background material. We recall, for context, but are not going to invoke, the classical Gilbert-Varshamov bound
$$r \ge 1-H(\delta).$$

\subsection{Polycyclic codes}
We say that a linear code $C$ of length $n$ over a field $F$  is {\bf polycyclic} if there exists a vector $\vc = (c_0,c_1,\dots,c_{n-1}) \in F^n$
such that for every $(a_0,a_1,\dots,a_{n-1} ) \in C$  we have
$$(0,a_0,a_1,\dots,a_{n-2}) + a_{n-1} (c_0,c_1,\dots,c_{n-1}) \in C.$$    We refer to $\vc$ as an {\bf associate vector}  of $C$. Note that
such a vector may be not unique.

To an associate vector $\vc$  we attach the polynomial $c(x) = c_0 + c_1x + c_2 x^2 + \dots + c_{n-1} x^{n-1}.$  Let $f(x) = x^n - c(x).$
It is shown in \cite{LPS} that polycyclic codes are ideals in $R_f=F[x] / \langle f(x) \rangle$ with the usual correspondence between vectors and polynomials

It is shown in \cite{LPS}, that a polycyclic code with the associate vector $\vc$ is left invariant by right multiplication of the matrix $D$ of the form:
\begin{equation}\label{eqnD} D = \left( \begin{array}{ccccc}
0&1&0& \dots& 0 \\
0& 0 & 1 & \dots & 0 \\
\vdots \\
0& 0 & 0 & \dots & 1 \\
c_0 & c_1 & c_2 & \dots & c_{n-1} \\
\end{array}\right ). \end{equation}

We call the endomorphism of $F^n$ with matrix $D^t$ the {\bf polyshift} associated with $\vc,$ and denote it by $T_{\vc}.$
The matrix $D^t$ is called in linear algebra the {\bf companion matrix} of $f(x)$ \cite[\S 7.1]{HK}.
If $\vc=(1,0,\dots,0)$ then $T_{\vc}$ is just the standard cyclic shift to the right. Sometimes, to avoid double indices, we will write $T$ for $T_{\vc}.$

\subsection{Double polycirculant codes}
A matrix $A$ of size $n \times n$ is {\bf polycirculant} for a polyshift $T_{\vc}$ if its rows are in succession $$\mathbf{a},
                                                                             T_{\vc}(\mathbf{a}),
                                                                             T_{\vc}^2(\mathbf{a}),
                                                                             \dots,
                                                                             T_{\vc}^{n-1}(\mathbf{a}). $$
                                                                             Such a matrix is uniquely determined by its first row and the associate vector $\vc.$
 If $\vc=(1,0,\dots,0)$ then $A$ is just a circulant matrix.
 A linear code $C$ of length $2n$ is said to be {\bf double polycirculant} (DP) if   its generator matrix $G$ is of the form
 $G=(I,A),$ where $I$ is the identity matrix of size $n \times n,$ and $A$ is a polycirculant matrix of the same size. If $\vc=(1,0,\dots,0)$, then $C$ is a (pure) double circulant code \cite{AOS}.

 {\bf Caveat:} We use $N$ for the length of the DP code, and $n$ for the size of $A$. Thus $N=2n$ in the whole paper.

\section{Duality results}

\subsection{Isoduality}
The following proposition is our main motivation to introduce double polycirculant codes. Throughout the paper the exponent $^t$ denotes transposition.
{\prop Let $A$ be a matrix satisfying $A^t=QAQ,$ with $Q$ a monomial matrix that satisfies $Q^2=I,$ where $I$ is identity of order $n.$ The code  $C=\langle(I,A)\rangle$  is an isodual code of length $2n.$}

\begin{proof}
The parity-check matrix of $C$ is then $H=(-A^t,I).$ Recall that $H$ spans $C^\perp.$ Using the hypothesis, we have $H\mathcal{Q}=(-QA,Q),$ where

$\mathcal{Q}=\left(\begin{matrix}Q & 0 \\0 & Q \end{matrix}\right).$ Hence $QH\mathcal{Q}=(-A,I),$ a matrix which spans a code equivalent to $C.$ The result follows.
\end{proof}

{\bf Remark 1:} In general it is known that any matrix $A,$ invertible over some field satisfies $A^t=uAu,$ with $u^2=I$ \cite{Go}. It is not known when $u$ can be a monomial matrix.

Now we exhibit a class of DP codes where this proposition applies.

{\thm \label{fonda} Given an associate vector $\mathbf{c}=(c_{0},c_{1},...,c_{n-1})$, and the vector $\mathbf{a}=(a_{1},a_{2},...,a_{n})$, define the polycirculant matrix $A$ by the equation $$A=\left(
                                                                            \begin{array}{c}
                                                                             \mathbf{a} \\
                                                                             T_{\vc}(\mathbf{a}) \\
                                                                             T_{\vc}^2(\mathbf{a})\\
                                                                             ... \\
                                                                             T_{\vc}^{n-1}(\mathbf{a}) \\
                                                                            \end{array}
                                                                        \right).$$
 The matrix $T_{\vc}$ is the companion matrix of some polynomial $f$.
If $f$ is of the form  $x^n-ax^m-b$ with $a,b \in F\setminus\{0\},$ then there is a monomial matrix $Q$ of order $2$ such that $AQ=QA^t.$ Namely, one can take $$Q_{i,j}=\left\{
\begin{matrix}
1,&\mbox{if} & i+j=m+1,\\
b, &\mbox{if} &i+j=n+m+1,\\
0,&\mbox{otherwise}.
\end{matrix}
\right.$$
In particular, under these hypotheses, the code $C=\langle(I,A)\rangle$ is isodual.
}

\begin{proof} We know that for any vector $\mathbf{c}=(c_{0},c_{1},...,c_{n-1})$, the matrix $$T_{\vc}=T=\left(
                                            \begin{matrix}
                                            0&0&\cdots&0&c_{0} \\
                                            1&0&\cdots&0&c_{1}\\
                                            0&1&\cdots&0&c_{2}\\
                                             \cdots&\cdots&\cdots&\cdots&\cdots \\
                                             0&0&\cdots&0&c_{n-2}\\
                                             0&0&\cdots&1&c_{n-1}\\
                                             \end{matrix}
                                              \right),$$
and it's the companion matrix of $f(x)=x^n-c_{n-1}x^{n-1}-...-c_{1}x-c_{0}$ over $\F_q.$ So, if $f(x)=x^n-ax^m-b,$ then $c_{0}=b,\,c_{m}=a$, $c_{1}=...=c_{m-1}=c_{m+1}=...=c_{n-1}=0$. If we let $E=AQ,$ $F=QA^{t},$ $E=(E_{i,j})_{n\times n},$ $F=(F_{i,j})_{n\times n}$, $A=(A_{i,j})_{n\times n},$ then we obtain $E_{i,j}=\sum\limits_{k=1}^{n}A_{i,k}Q_{k,j}$, $F_{i,j}=\sum\limits_{k=1}^{n}Q_{i,k}A_{j,k}$, and, by the definition of $Q$, we can obtain $$E_{i,j}=\left\{
\begin{matrix}
A_{i,m+1-j},&1\leq j\leq m;\\
bA_{i,n+m+1-j},&m<j\leq n;
\end{matrix}
\right.
F_{i,j}=\left\{
\begin{matrix}
A_{j,m+1-i},&1\leq i\leq m;\\
bA_{j,n+m+1-i},&m<i\leq n.
\end{matrix}
\right.$$ Now we have to prove $E_{i,j}=F_{i,j}$ for $i,j=1,2,...,n$. \\
  Let $T_{j}$ be the $j$-th row of the matrix $T$. By the definition of $A$, we can get
  $$(A_{i,1},A_{i,2},...,A_{i,n})=T_c(A_{i-1,1},A_{i-1,2},...,A_{i-1,n})=(A_{i-1,1},A_{i-1,2},...,A_{i-1,n})T^t.$$
  Then
  $$A_{i,j}=(A_{i-1,1},A_{i-1,2},...,A_{i-1,n})T_{j}^t=\left\{
                                      \begin{matrix}
                                      bA_{i-1,n},&j=1;\\
                                      A_{i-1,m}+aA_{i-1,n},&j=m+1;\\
                                      A_{i-1,j-1},&otherwise.
                                      \end{matrix}
                                       \right.$$
                                       We distinguish four cases depending on the relative positions of $i$ and $j$ with respect to $m.$
                                       \begin{enumerate}
\item [(i)] \underline{\bf If $1\leq i\leq m$ and $1\leq j\leq m$,} then we have  $E_{i,j}=A_{i,m+1-j}$ and $F_{i,j}=A_{j,m+1-i}$.
In this case, $1\leq m+1-i\leq m$ and $1\leq m+1-j\leq m$. If $i\geq j$, when $j=m$, we can get $i=j=m$, $E_{i,j}=A_{m,i}=F_{i,j}$; when $1\leq j<m$, we can get $A_{i,m+1-j}=A_{i-(i-j),m+1-j-(i-j)}=A_{j,m+1-i}.$ If $i<j$, we can get $A_{j,m+1-i}=A_{j-(j-i),m+1-i-(j-i)}=A_{i,m+1-j}.$ It means that $E_{i,j}= F_{i,j}$.
\item [(ii)] \underline{\bf If $1\leq i\leq m$ and $m<j\leq n$,} then we have $E_{i,j}=bA_{i,n+m+1-j}$ and $F_{i,j}=A_{j,m+1-i}$.
In this case $1\leq m+1-i\leq m$ and $m+1\leq n+m+1-j\leq n$. Furthermore, $m+1-i\leq j-i$, implying $A_{j,m+1-i}=A_{j-(m-i),m+1-i-(m-i)}=A_{i+j-m,1}=bA_{i+j-m-1,n}=bA_{i,n+m+1-j},$ thus $E_{i,j}= F_{i,j}$.
\item [(iii)] \underline{\bf If $m<i\leq n$ and $1\leq j\leq m$,} then we have $E_{i,j}=A_{i,m+1-j}$ and $F_{i,j}=bA_{j,n+m+1-i}$.
The proof follows as in case (ii).
\item [(iv)] \underline{\bf If $m<i\leq n$ and $m<j\leq n$,} then we have $E_{i,j}=bA_{i,n+m+1-j}$ and $F_{i,j}=bA_{j,n+m+1-i}$. In this case $m+1\leq n+m+1-i\leq n$ and $m+1\leq n+m+1-j\leq n$. The proof follows as in case (i).
\end{enumerate}
The last assertion follows by Proposition 1.
This completes the proof.\end{proof}
\subsection{Self-duality criterion when $q=2$}
In this subsection, and the next one, we work over $\F_2.$
Given the vector $\mathbf{c}=(c_{1},c_{2},...,c_{n})$ where $c_{1}=c_{m}=1,c_{i}=0,i\neq 1,m$, and the first row $\mathbf{a}=(a_{1},a_{2},\cdots,a_{n})$ define the matrix $A=\left(
                                                                            \begin{array}{c}
                                                                             \mathbf{a} \\
                                                                             T_{\vc}(\mathbf{a}) \\
                                                                             T_{\vc}^2(\mathbf{a})\\
                                                                             ... \\
                                                                             T_{\vc}^{n-1}(\mathbf{a}) \\
                                                                            \end{array}
                                                                        \right)$.
When $\vc$ satisfies the above conditions, $T_{\vc}$ is the companion matrix of the trinomial $f(x)=x^n-x^{m-1}-1.$
{\thm With the above notation, the code $C=\langle(I,A)\rangle$ where $I$ is identity matrix of order $n$ is a binary self-dual code  if and only if $A=I.$
In other words, a  binary DP code with a trinomial induced polyshift is a self-dual code iff it is equivalent to a direct sum of repetition codes of length $2.$
}

\begin{proof}
The code $C=\langle(I,A)\rangle$ is a self-dual code iff $AA^{t}=-I=I$ over $\F_{2}$, where $A^{t}$ is the transpose of $A$. Now we need to prove $AA^t=I$ iff $A=I$.
On the one hand, let $B=AA^{t}$, $B=(B_{i,j})_{n\times n}$,
$A=(A_{i,j})_{n\times n}$, then $B_{i,j}=\sum\limits_{k=1}^{n}A_{i,k}A_{j,k}$.\\

With the definition of $A$, we can get the $i$-th row of $A$ is
\begin{align*}
(A_{i,1},A_{i,2},...,A_{i,n})&=(0,A_{i-1,1},A_{i-1,2},...,A_{i-1,n-1})+A_{i-1,n}\vc\\
                             &=(0,A_{i-1,1},A_{i-1,2},...,A_{i-1,n-1})+A_{i-1,n}(1,0,...,0,1,0,...,0),
\end{align*}
then \begin{align*}
A_{i,1}^2+A_{i,2}^2+...+A_{i,n}^2&=A_{i-1,1}^2+A_{i-1,2}^2+...+A_{i-1,n-1}^2+A_{i-1,n}^2+A_{i-1,n}^2,\\
\sum\limits_{k=1}^{n}A_{i,k}&=\sum\limits_{k=1}^{n}A_{i-1,k}+A_{i-1,n}.
\end{align*}
With the definition of $B$, we have
\begin{align*}
B_{1,1}&=\sum\limits_{k=1}^{n}A_{1,k}^2=\sum\limits_{k=1}^{n}A_{1,k}=a_{1}+a_{2}+...+a_{n-1}+a_{n};\\
B_{2,2}&=\sum\limits_{k=1}^{n}A_{2,k}^2=\sum\limits_{k=1}^{n}A_{2,k}=B_{1,1}+A_{1,n};\\
B_{3,3}&=\sum\limits_{k=1}^{n}A_{3,k}^2=\sum\limits_{k=1}^{n}A_{3,k}=B_{2,2}+A_{2,n};\\
&...\\
B_{n-1,n-1}&=\sum\limits_{k=1}^{n}A_{n-1,k}^2=\sum\limits_{k=1}^{n}A_{n-1,k}=B_{n-2,n-2}+A_{n-2,n};\\
B_{n,n}&=\sum\limits_{k=1}^{n}A_{n,k}^2=\sum\limits_{k=1}^{n}A_{n,k}=B_{n-1,n-1}+A_{n-1,n}.
\end{align*}
If $B=I$, $B_{1,1}=B_{2,2}=...=B_{n,n}=1$, then we can get $A_{1,n}=A_{2,n}=...=A_{n-1,n}=0$.
Let $T_{j}$ be the $j$-th row of matrix $T.$  Since the vector $\mathbf{c}=(c_1,c_2,...,c_n)$ with $c_1=c_m=1$, $c_{i}=0$, $i\neq1,m$, we have $$T_{j}=\left\{
                                      \begin{matrix}
                                      e_{n},&j=1;\\
                                      e_{m-1}+e_{n},&j=m;\\
                                      e_{j-1},&otherwise.
                                      \end{matrix}
                                       \right.$$
  where $e_{i},i=1,2,...,n$ is the canonical basis of $F^n,$ defined by $(e_{i})=\delta_{ij},i,j=1,2,...,n$. Thus, by computations similar to those in the proof of Theorem 1, we have
  $$A_{i,j}=T_{j}\left(
                       \begin{matrix}
                        A_{i-1,1}\\
                        A_{i-1,2}\\
                        ...\\
                        A_{i-1,n}
                       \end{matrix}
                        \right)=\left\{
                                      \begin{matrix}
                                      A_{i-1,n},&j=1; \ \ \ \ \ \ \ \ \ \ \ \ \ (1)\\
                                      A_{i-1,m-1}+A_{i-1,n},&j=m; \ \ \ \ \ \ \ \ \ \ \ \ (2)\\
                                      A_{i-1,j-1},&otherwise.\ \ \ \ \ \  \ \ (3)
                                      \end{matrix}
                                       \right.$$
  If $1\leq t\leq n-m$, by $(3)$ we can get
  $$A_{t,n}=A_{t-(t-1),n-(t-1)}=A_{1,n-t+1}=a_{n-t+1}.$$
  So for each $m+1 \le j\leq n$ we have $$a_j=A_{n-j+1,n}.$$
  If $n-m+1\leq t\leq n$, by $(3)$ we can get
  $$A_{t,n}=A_{t-(n-m),n-(n-m)}=A_{t+m-n,m}.$$
  So for each $1 \le i\leq m$ we have $$A_{i,m}=A_{n-m+i,n}.$$
  By applying $(2)$ and then $(3)$, it follows that for each $1<i\leq m$ we have
  $$A_{i,m}=A_{i-1,m-1}+A_{i-1,n}=A_{1,m-i+1}+A_{i-1,n}=a_{m-i+1}+A_{i-1,n}.$$
  Then for $1\leq j<m$ we have $$a_j=A_{m-j+1,m}-A_{m-j,n}=A_{n-j+1,n}-A_{m-j,n}.$$
  What's more, $a_m=A_{1,m}=A_{n-m+1,n}.$

  Because $A_{1,n}=A_{2,n}=...=A_{n-1,n}=0$, so $a_n=a_{n-1}=...=a_2=0$; further, $B_{1,1}=1$, yielding $a_1=1$. Then with ${\bf a}=(a_1,a_2,...,a_n)=(1,0,...,0)$, we check from its definition the matrix $A=\left(
                                            \begin{matrix}
                                            1&0&\cdots&0&0 \\
                                            0&1&\cdots&0&0\\
                                             \cdots&\cdots&\cdots&\cdots&\cdots \\
                                             0&0&\cdots&1&0\\
                                             0&0&\cdots&0&1\\
                                             \end{matrix}
                                              \right)=I.$\\

  On the other hand, if $A=I$, it is immediate that $AA^{t}=I$.
\end{proof}
\subsection{Evenness}
An important class of binary isodual codes is the class of even isodual codes that is those isodual codes all weights of which are even \cite{FGPH,KP}. The next result shows that over $\F_2$ many isodual codes coming from our construction are odd.
{\thm Let $A$ be a polycirculant matrix of size $n\times n,$ for a polyshift $T_{\vc}.$ If $\vc$ is an even weight vector,
then the binary code $C=\langle(I,A)\rangle$ is an even code if and only if the first row of $A$ is $\mathbf{a}=(1,0,\cdots,0)$.}

 \begin{proof}
 If $C=\langle(I,A)\rangle$ is an even code, then each row of $A$ has an odd weight. \\
Firstly, suppose the weight of $\mathbf{a}$ is $W(\mathbf{a})=1$, namely, $\mathbf{a}=(a_{1},a_{2},\cdots,a_{n})$ with only $a_{m}=1$. By the definition of $A$, we can get the first $n-m+1$ rows of $A$ as:
\begin{align*}
             \mathbf{a}&=(0,...,0,1,0,0,0,...,0,0);\\
      T_{\vc}(\mathbf{a})&=(0,...,0,0,1,0,0,...,0,0);\\
  T_{\vc}^{2}(\mathbf{a})&=(0,...,0,0,0,1,0,...,0,0);\\
  &...\\
T_{\vc}^{n-m}(\mathbf{a})&=(0,...,0,0,0,0,0,...,0,1).
\end{align*}
Then the $(n-m+2)$-th row is $$T_{\vc}^{n-m+1}(\textbf{a})=(0,0,...,0,0)+1\cdot \mathbf{c},$$
and the weight of this row is $W(T_{\vc}^{n-m+1}(\textbf{a}))=W(\mathbf{c})\equiv 0 \pmod{2}.$ So in this case, if we expect that the weight of each row of $A$ is odd, it requires $n-m=n-1$, then $m=1$ and $\mathbf{a}=(1,0,\cdots,0)$.

More generally, suppose that the weight of $\mathbf{a}$ is odd, with rightmost nonzero $a_i$ for $i=t.$ If we let $W(\mathbf{a})=2k+1,~k\geq0$, and $\textbf{a}=(0,...,1,0,...,1,0,...,1,0,...,0)$ with rightmost nonzero $a_t=1$, reasoning as before, by the definition of $A$, we can get the first $n-t+1$ rows of $A$ as:
\begin{align*}
             \mathbf{a}&=(0,...,1,0,...,1,0,...,1,0,0,0,...,0);\\
      T_{\vc}(\mathbf{a})&=(0,0,...,1,0,...,1,0,...,1,0,0,...,0);\\
  T_{\vc}^{2}(\mathbf{a})&=(0,0,0,...,1,0,...,1,0,...,1,0,...,0);\\
  &...\\
T_{\vc}^{n-t}(\mathbf{a})&=(0,...,0,0,0,0,...,1,0,...,1,0,...,1).
\end{align*}

Then the $(n-t+2)$-th row is $$T_{\vc}^{n-t+1}(\textbf{a})=(0,0,...,0,0,0,0,...,1,0,...1,0,...,0)+1\cdot \mathbf{c},$$
and the weight of this row is $W(T_{\vc}^{n-t+1}(\textbf{a}))=2k+W(\textbf{c})$. Because $W(\mathbf{c})\equiv 0 \pmod{2},$ in this case, if we expect that the weight of each row of $A$ is odd, it requires $n-t=n-1$, then $t=1$. It means that $a_1=1$ is the rightmost nonzero element of $\textbf{a}$. Therefore, $\textbf{a}=(1,0,...,0)$.

On the other hand, if the vector $\mathbf{a}=(1,0,...,0)$, then we can check from the definition of $A$ that $A=I$.
It is immediate to check that $C=\langle(I,A)\rangle=\langle(I,I)\rangle$ is an even code.
\end{proof}

{\bf Remark 2:} When $\textbf{c}$ has odd weight, we can get that the weight of each row of A is odd if and only if the weight of \textbf{a} is odd by a similar argument as in the proof of Theorem 3. Namely, the binary code $C=\langle(I,A)\rangle$ is even if and only if \textbf{a} has odd weight. This extension is left to the reader.

\section{Asymptotics}
We prepare for the main result of the section by the following lemma. Recall that the {\bf Cyclotomic polynomial} of index $m$ denoted here by $Q_m(.),$ is the $\Z$-polynomial with roots all the elements of order $m$ in the algebraic closure of $\Q$ \cite{LN}.

{\lem\label{irr} For any positive integer $n$, the trinomial $H_n(x)=x^{2.3^n}+x^{3^n}+1$ is irreducible over $\F_2.$ }

\begin{proof}
The cyclotomic polynomial $Q_m(x)$ is irreducible over $\F_2$  iff its degree $\phi(m)$ is equal to  the order of $2 \pmod{m}$, that is the minimal positive integer $k$ such that $2^k\equiv 1\pmod{m}$ \cite[Th. 2.47 (ii)]{LN}. Let $m=3^{n+1}.$ Thus, by \cite[Ex. 2.46]{LN}, we know that  $Q_{m}(x)=H_n(x)=x^{2.3^n}+x^{3^n}+1$, and well-known properties of Euler totient function show that $\phi(m)=2.3^{n}$.  By \cite[Thm 1]{cprpp}, because $2$ is a primitive root $\pmod{3},$ it  is a primitive root $\pmod{3^n}$ for all $n\ge 1.$ Thus $Q_{m}(x)=H_n(x)$ is irreducible.
\end{proof}
The number of DP codes of length $2n$ is important to count.
{\prop\label{omega} For a given polyshift, the number of DP codes of length $2n$ over $\F_q$ is $q^n.$}

The easy proof is omitted. We prepare for the proof of the next theorem by a lemma from linear algebra.

{\lem Let $0\neq \mathbf{w} \in \F_q^n.$ Let $R$ be a matrix of size $n\times n$ over $\F_q$ with an irreducible characteristic polynomial.
The matrix with successive rows $\mathbf{w},R\mathbf{w},\dots,R^{n-1}\mathbf{w}$ is nonsingular.}

\begin{proof} The result is immediate from \cite[Chap. 7, Theorem 1 (ii)]{HK}. To be self-contained we give a proof.
Consider the $\F_q$ vector space $$ V=\{R^i\mathbf{w}\mid i=0,1,\cdots,n\}.$$ This vector space is invariant by $R.$ Let $h$ (resp.$\chi$) denote  the minimal (resp. characteristic) polynomial of $R$ restricted to $V.$ Let $g$ be the characteristic polynomial of $R$ on the whole space $\F_q^n.$ Then by the lemma of \cite[p.200]{HK}, the polynomial $\chi$ divides $g.$ By Cayley-Hamilton theorem, $h$ divides $\chi.$ Hence $h$ divides $g.$ By \cite[\S 7, Theorem 1]{HK}
the degree of $h$ equals the dimension of $V.$ Since $g$ is irreducible, this means that $h=g$ and $V=\F_q^n.$ The result follows.
\end{proof}

{\thm \label{lambda} Let $\mathbf{u},\mathbf{v} \in \F_q^n.$ For a given polyshift,
the matrix of which is the companion matrix of an irreducible trinomial,  the number of DPs of length $2n$ over $\F_q$ that contains the vector $(\mathbf{u},\mathbf{v}) \neq 0$ is at most one.}

\begin{proof}
We need to solve $A^t\mathbf{u}^t=\mathbf{v}^t,$ when $A^t=QAQ,$ and $Q$ is as in Theorem \ref{fonda}.  Letting $\mathbf{u'}=Q\mathbf{u}^t$ and $\mathbf{v}'=Q\mathbf{v}^t,$ $\mathbf{v}'=(v_1',v_2',...,v_n')^t$, we obtain the system of $n$ equations $$v'_1=\mathbf{a}\cdot \mathbf{u'},\dots,v'_i=T_{\vc}^{i-1}(\mathbf{a})\cdot \mathbf{u}',\dots,v'_n=T_{\vc}^{n-1}(\mathbf{a})\cdot \mathbf{u'}.$$ By transposition,
we obtain a system in the $a_i$'s whose matrix has successive columns $$\mathbf{u'},\dots,T_{\vc}^t\mathbf{u'},\dots,(T_{\vc}^t)^{n-1}\mathbf{u'}.$$ By the lemma this matrix is non singular. The result follows.

\end{proof}

We are now in a position to state and prove the main result of this section. The proof technique is the classical method of expurgated random coding.

{\thm  For all $0<\delta < H^{-1}(\frac{1}{2}),$ there are sequences of binary DP codes of relative distance $\delta$ with a polyshift
whose matrix is the companion of an irreducible trinomial.}

\begin{proof}
By Lemma \ref{irr}, there are infinitely many irreducible trinomials over $\F_2.$ For a given $n,$ by Proposition \ref{omega} there are
$\Omega_n= 2^n$ DP codes of length $2n$ w.r.t. a polyshift whose matrix is the companion matrix of a given irreducible trinomial. We will require the following entropic estimate.
The total number of binary vectors of length $2n,$ and Hamming weight $<d_n=\lfloor 2\delta n\rfloor$, $V_n$ say, is at most
\begin{equation}\label{ms}
  V_n\le{2^{2nH(\delta)}} 
\end{equation}
by \cite[Chap. 10, Cor. 9 ]{MS}.

By Theorem \ref{lambda}, a nonzero vector $(\mathbf{u},\mathbf{v})$ with weight $<d_n$ can be contained in at most one such code.
If \begin{equation}\label{funda}
    \Omega_n>V_n,
   \end{equation}
 then there is at least one such DP code of length $2n$ with minimum distance $\ge d_n.$
 (Note that it is essential that inequality (\ref{funda}) be strict to derive that conclusion).

 By (\ref{ms}) we see that inequality (\ref{funda})
will hold for $n$ large enough if $$ {2^{2nH(\delta)}}=o(2^n),$$ which will hold in particular if
 $H(\delta)<1/2.$
\end{proof}

{\bf Remark:} Thus this theorem means that for every $\epsilon>0,$ there are sequences of DP codes with a relative distance $> H^{-1}(\frac{1}{2})-\epsilon.$
Note, for sake of comparison, that the quantity $H^{-1}(\frac{1}{2})$ is the Gilbert-Varshamov bound on the relative distance of linear binary codes of rate $1/2.$.

\section{Numerics}

In Tables 1-2 and 4-5, for, respectively, $q=2,3,5,7$ we denote by
\begin{itemize}
\item $d_F(q,2n)$ the highest minimum weight of formally self-dual codes over $\F_q$ as per \cite{BH,DGH0,DGH},

\item $d_F^\ast(q,2n)$ the highest minimum weight of FSD DP codes constructed over $\F_q.$
\end{itemize}
We put a star exponent on the entry $d_F^\ast(q,2n)$ whenever $d_F^\ast(q,2n)=d_F(q,2n).$
In Table 3 we denote by $d_{fsdao}(4,2n)$ the  highest minimum weight of formally self-dual additive odd codes over $\F_4$ (\cite{HaK}); and by $d_{fsdao}^\ast(4,2n)$ the highest minimum weight of FSD DP codes that we can find over $\F_4$.
We put a star exponent on the entry $d_{fsdao}(4,2n)^\ast$ whenever $d_{fsdao}(4,2n)^\ast=d_{fsdao}(4,2n).$
We constructed a large number of random DP codes with trinomial polyshifts as in Theorem 1, and the Tables collect the best found.
All binary codes constructed are odd. All the DP codes constructed in this section are FSD codes by Theorem 1.
All computations were performed in Magma \cite{M}.
\begin{table}
\centering
\begin{center}
\small {Table $1$: The Highest Minimum Weight for $\F_2$} \\
\end{center}
  \begin{tabular}{ccc}
   \toprule
    Length $2n$ &$d_F^\ast(2,2n)$&$d_F(2,2n)$\\
   \midrule
     4&2$^\ast$&2 \\
     6&3$^\ast$&3 \\
     8&3$^\ast$&3 \\
     10&3&4 \\
     12&4$^\ast$&4 \\
     14&4$^\ast$&4\\
     16&5$^\ast$&5\\
     18&5$^\ast$&5\\
     20&5&6\\
     22&6&7\\
     24&6&7\\
     26&6&7\\
     28&6&7\\
     30&7$^\ast$&7 or 8\\
     32&7&8\\
     34&7&8\\
     36&8$^\ast$&8\\
     38&8$^\ast$&8 or 9\\
     40&8&9 or 10\\
   \bottomrule
  \end{tabular}
\end{table}
\pagebreak

\begin{table}
\centering
\begin{center}
\small {Table $2$: The Highest Minimum Weight for $\F_3$} \\
\end{center}
  \begin{tabular}{ccc}
   \toprule
    Length $2n$ &$d_F^\ast(3,2n)$&$d_F(3,2n)$\\
   \midrule
     4&3$^\ast$&3 \\
     6&3$^\ast$&3 \\
     8&4$^\ast$&4 \\
     10&4&5 \\
     12&5&6 \\
     14&5&6\\
     16&6$^\ast$&6\\
     18&6$^\ast$&6\\
     20&6&7\\
     22&7&8\\
     24&8&9\\
     26&8$^\ast$&8 or 9\\
     28&8&9 or 10\\
     30&8&9, 10 or 11\\
   \bottomrule
  \end{tabular}
\end{table}
\pagebreak
\begin{table}
\centering
\begin{center}
\small {Table $3$: The Highest Minimum Weight for $\F_4$} \\
\end{center}
  \begin{tabular}{ccc}
   \toprule
    Length $2n$ &$d_{fsdao}^\ast(4,2n)$&$d_{fsdao}(4,2n)$\\
   \midrule
     4&3$^{\ast}$&$3$ \\
     6&3$^{\ast}$&$3 $\\
     8&4$^{\ast}$&$4$ \\
     10&5$^{\ast}$&$5$ \\
     12&5&$6$ \\
     14&6$^{\ast}$&6 or $7$\\
   \bottomrule
  \end{tabular}
\end{table}
\begin{table}
\centering
\begin{center}
\small {Table $4$: The Highest Minimum Weight for $\F_5$} \\
\end{center}
  \begin{tabular}{ccc}
   \toprule
    Length $2n$ &$d_F^\ast(5,2n)$&$d_F(5,2n)$\\
   \midrule
     4&3$^\ast$&3 \\
     6&4$^\ast$&4 \\
     8&4$^\ast$&4 \\
     10&5$^\ast$&5 \\
     12&6$^\ast$&6 \\
     14&6$^\ast$&6\\
     16&7$^\ast$&7\\
     18&7$^\ast$&7 or 8\\
     20&7&8 or 9\\
     22&8$^\ast$&8, 9 or 10\\
     24&8&9 or 10\\
   \bottomrule
  \end{tabular}
\end{table}
\pagebreak

\begin{table}
\centering
\begin{center}
\small {Table $5$: The Highest Minimum Weight for $\F_7$} \\
\end{center}
  \begin{tabular}{ccc}
   \toprule
    Length $2n$ &$d_F^\ast(7,2n)$&$d_F(7,2n)$\\
   \midrule
     4&3$^\ast$&3 \\
     6&4$^\ast$&4 \\
     8&4&5 \\
     10&5$^\ast$&5 \\
     12&6$^\ast$&6 \\
     14&6&7\\
     16&7$^\ast$&7 or 8\\
     18&7&8 or 9\\
     20&8&9 or 10\\
     22&8&9, 10 or 11\\
     24&9&10, 11 or 12\\
   \bottomrule
  \end{tabular}
\end{table}

\section{Conclusion and open problems}
In this work we have introduced a new class of QPC codes of index two: double polycirculant codes. These codes are the natural generalization of double circulant codes when going from standard shift to the polyshift of polycylic codes. When the matrix of the polyshift is the companion matrix of a trinomial of the form $x^n+ax^m+b,$ these codes are isodual, and in particular formally self-dual. In short lengths, their parameters are optimal or quasi-optimal amongst FSD codes. More importantly, when $q=2,$ we could show that they are asymptotically good. 

Many open problems remain. Characterizing the matrices over finite fields that are monomially  equivalent to their transpose could lead to new constructions of isodual codes. At a more structural level, it might be possible to derive isoduality for a larger class of polyshifts. The characterization of evenness is only done for half the associate vectors. It would be worth constructing even FSD codes over $\F_2,$ the first studied class of FSD codes \cite{FGPH,KP}, by DP codes, or to prove they do not exist. Eventually, DP codes over finite rings is
a wide open area, which is well worth investigating, in view of the double circulant codes over rings of \cite{BGH}.

\end{document}